# Common Mistakes when Applying Computational Intelligence and Machine Learning to Stock Market modelling.

E. Hurwitz
University of Johannesburg
Johannesburg, South Africa
hurwitze@gmail.com

T. Marwala.
University of Johannesburg
Johannesburg, South Africa
tmarwala@gmail.com

*Abstract*—For a number of reasons, computational intelligence and machine learning methods have been largely dismissed by the professional community. The reasons for this are numerous and varied, but inevitably amongst the reasons given is that the systems designed often do not perform as expected by their designers. The reasons for this lack of performance is a direct result of mistakes that are commonly seen in market-prediction systems. This paper examines some of the more common mistakes, namely dataset insufficiency; inappropriate scaling; time-series tracking; inappropriate target quantification and inappropriate measures of performance. The rationale that leads to each of these mistakes is examined, as well as the nature of the errors they introduce to the analysis / design. Alternative ways of performing each task are also recommended in order to avoid perpetuating these mistakes, and hopefully to aid in clearing the way for the use of these powerful techniques in industry.

Keywords; Computational intelligence, machine learning, stock market, equities, automated stock tradin, mistakes.

## I. INTRODUCTION

The promise of computational intelligence and machine learning for application to equity price modeling has been obvious to almost everyone to have even dabbled in the fields since their inception. It is also unfortunately true that attempts to make such applications accepted within the asset management industry have been largely unsuccessful, and not without reason. Common mistakes have continued to provide papers and theories that have passed cursory academic scrutiny, but have shown to be ineffectual when applied to actual markets [1]. Within this paper, a number of common mistakes will be interrogated, illustrating how they usually slip past cursory academic scrutiny, and then showing how they fail when applied to actual market data and why. The hope is that with a spotlight lit upon common failings, researchers can avoid these pitfalls and advance the uses of these powerful techniques within the financial modeling arena. While many of these errors overlap in individual implementations, this paper endeavours to tackle them on an individual basis in order to more easily avoid these mistakes in the future.

## II. METHODOLOGY

In order to illustrate these common mistakes, each will be examined in the following manner:

- The action will be illustrated, its usual reasoning explained.
- The mechanism of how the action appears valid will be examined.
- The flaws in the action will be examined, elaborating on the difference between the apparent expectations and the actual results.
- The alternative, correct approach is explained and motivated.

These steps will illustrate conclusively how the highlighted actions are in fact mistakes that in almost all cases render their underlying analysis invalid. Each individual mistake will be examined in isolation. It is assumed that all techniques are intended for trading purposes, and as such to be applied to unseen data.

## III. INSUFFICIENT DATASETS

Typical trading strategies are complex systems, generally involving a cycle of prediction, evaluation, feedback, and recalibration when being designed, as depicted in Fig 1. (the representation does not include a feedback loop as many predictive systems are content with what amounts to feedforward control). The cycle itself has a feedback element, but not necessarily the predictive method.

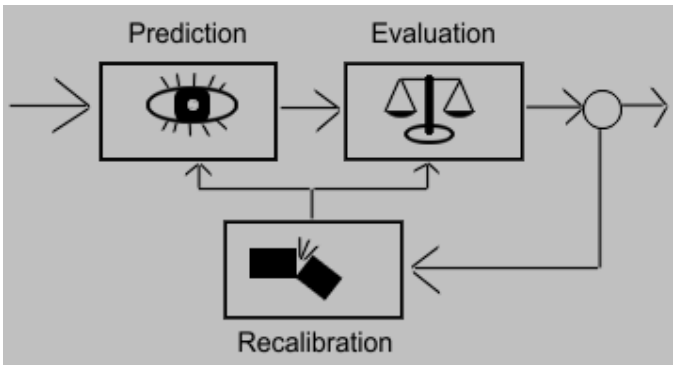

Figure 1. Trading system training cycle

This cycle involves the designer predicting price movements, then evaluating trades based on the price movements. The reason for recalibration feedback going into both Prediction and Evaluation mechanisms is that in many more complex system the actual trading based on the predictions will be updated, as well as the predictions themselves. As a result of this cycle, it becomes necessary to have a truly unseen set of data to evaluate performance. Too often trading systems are developed [2] [11] [12] [14] [17] that merely utilise a training data set and a verification data set. This may often seem sufficient, as neither the predictive system nor the evaluation system ever see the second set of data during the optimisation phase. This is however insufficient, as the calibrating of the training system is done with the results of the verification set taken into account, requiring a completely unseen validation set to be used in order to validate if the system is truly generalising. Omitting this crucial step can allow a system to result that has merely been tweaked by the designer to fit the specific data, without actually being able to function correctly in a general setting. The results will of course look good to the designer, as the system has been calibrated specifically to fit the data being used, even while the designer has not intended this to be the case. In the case that the trading mechanism is revisited after the validation set has been utilised, a new set of unseen validation data must be obtained. The easy rule of thumb is that if the trading system is altered N times, then the designer needs N+1 data sets.

## IV. INAPPROPRIATE SCALING

This error is typified by representing the typically large target values of the predicted variable (in most cases, the actual stock price) as its actual value [3], instead of scaling the data to some appropriate level within (or near) the range of the training data. The common reasoning is to provide an accurate view of the actual target values [3].

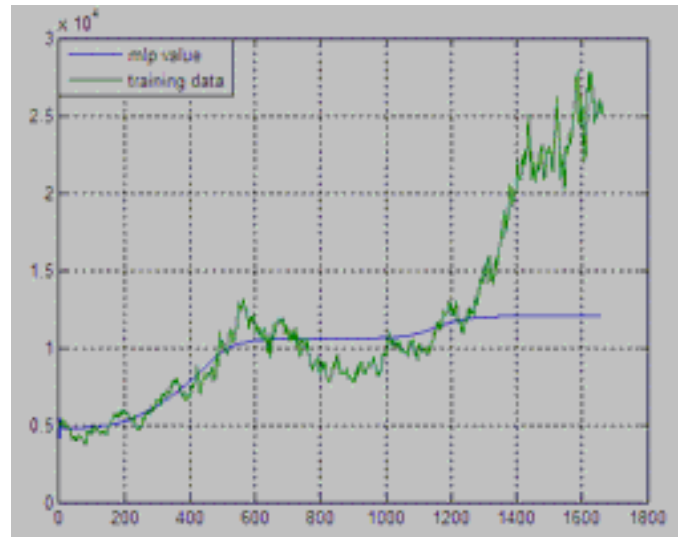

Figure 2. Unscaled Predictive System

An example of this can be viewed above in Fig 2. One can see how the actual data and the predicted data appear to close in the initial prediction, but struggles to reach the higher values in the range for the later values. The actual errors in this initial prediction will of course be very low, being dwarfed by the scale of the data, despite their being quite obviously useless for any task that requires the prediction. In fact, as elaborated on in section VII, the errors inherent in the system are often hidden through using inappropriate measures of accuracy / performance. It is in fact far more dangerous to make this error if the target data is reasonably bounded, as the obvious lack of fit seen in the latter half of Fig 2 will not be evident, and what is in fact a useless prediction can easily be mistaken for a performing predictive system, and leads to all the dangers inherent in trading upon poor information.

The reason for this discrepancy is the high quantitative value of the predictive results, which give a low registered error for what is in effect a large trading error. Consider a prediction for a given input-output set, with the correct value being 2050, and the system's predicted value being 2030. The actual error in RMS terms is tiny, while the effect on predictions is actually quite high, considering a daily expected fluctuation of approximately 30 cents, which explains why an error that appears so tiny is in fact more than large enough to render a predictive system useless. The obvious recommendation is to first pre-process the data, and as part of that process to scale the data. Depending on the nature of the historical share price fluctuations, a scaling factor of anything from the maximum historical price recorded to a fractional amount larger than the maximum historical price (allowing for higher future prices than previously recorded, although the system is unlikely to generalise outside of the trained region) can be used. Simply divide the target training data by this divisor, and one ensues that your target data will lie in the range of [0 1] (or slightly below one if the fractional approach is followed), which handles the scaling errors with aplomb. It is also recommended, although not strictly speaking relevant to this particular mistake, that one also scales input data for ease of training and convergence.

## V. TIME-SERIES TRACKING

In this next case, the old adage of "if it looks too good to be true, it probably is" is completely true. Results similar to those illustrated below in Fig 3 are seen in often when performing time-series analysis on a share price [8] [15] [16].

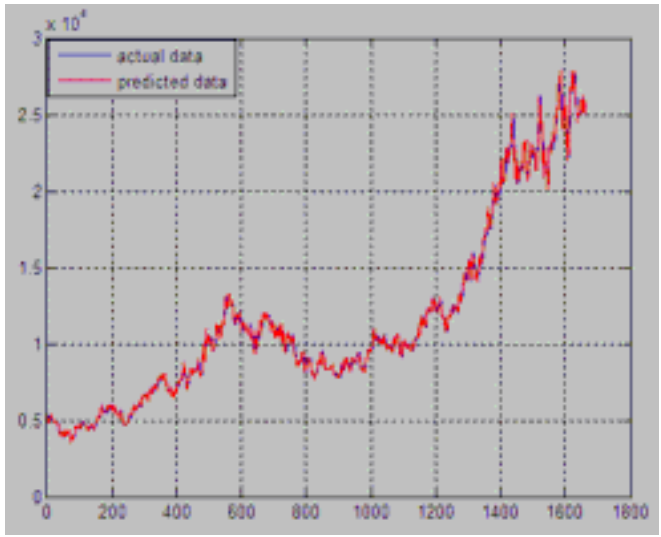

Figure 3. Too good to be true

So is it in fact too good to be true? If it is not, then all trading has become a very simple task. Unsurprisingly, the results depicted above are in fact too good to be true, and all it takes to expose the lie is some closer examination of the results, such as a zoomed in portion of the above graph, depicted in Fig 4. After a quick glance at Fig 4, the error should become immediately apparent – the system is doing little more than predict the previous day's price, which of course satisfies the error minimisation function's needs.

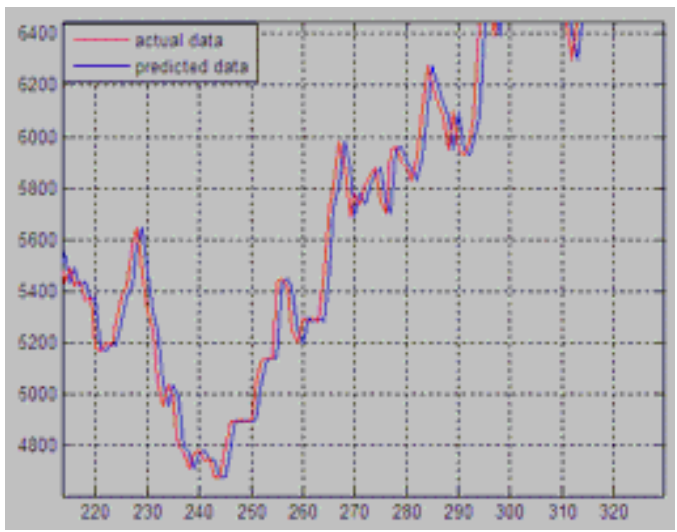

Figure 4. The lie exposed

This error comes about when the designer attempts to do an exact price prediction, based upon the time-series data of historical prices of the self-same share. The results of course look amazing at first glance, and one can only imagine euphoria and the prospect of easy riches prevents the researcher of digging a little deeper. Unfortunately any trading based upon such a system is completely useless, as it cannot ever predict an accurate price movement, unless by some absurd chance every single day's new close is the same as that of the day before. Should this error occur, it is then necessary for the designer to rethink the input-output pairs for the system to learn from. This type of error can also come about from an overly-focused mindset: A researcher looking at the problem as "what will tomorrow's price be?" can easily fall into the mistake just described, as he has not taken into account the desired application of his system. A more appropriate question would be "what will tomorrow's price be, in order that I can trade on it at a profit?". In this second case, the researcher notes that it is in order to trade that the prediction is being done, and so the usefulness of the prediction is of more importance than the raw error-value itself. In order to avoid this sort of error, especially when performing time-series analysis, it is recommended that the input and output data used in training the system not both be prices, or for that matter any value likely to be of a highly similar nature. It is always worth considering, when designing any predictive training system, "what exactly will the system see while it is learning?". Asking this question can head off many mistakes before they ever become manifest.

## VI. INAPPROPRIATE TARGET QUANTIFYING

As seen in section IV, trying to predict an exact day's price can lead to seemingly small errors that are in fact quite significant. Even when scaled appropriately, the prediction of an exact day's price can prove highly problematic, as illustrated in Fig 5.

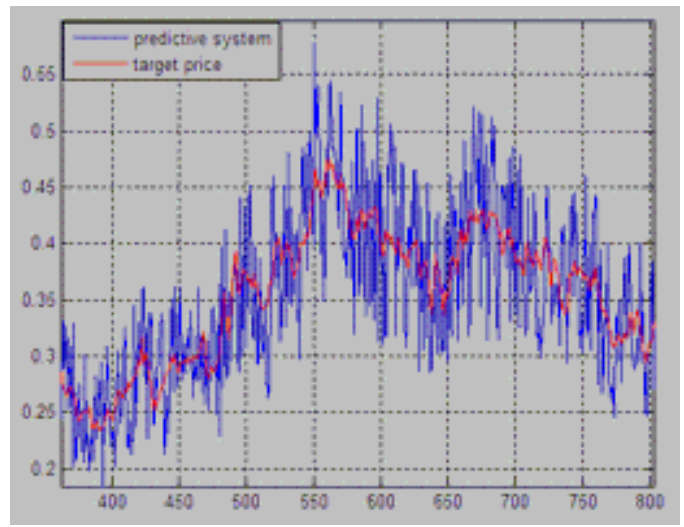

Figure 5. Scaled but poor prediction

While the prediction may look good, it is in fact badly flawed. Trading on the above prediction will lose money rapidly, despite what appears to be a high level of accuracy. This is because the system is not concerned with the daily direction of the price movements, but only with its proximity to target value. When trading, the direction of price movement is actually far more important than the precise amount, and this distinction is critical if a trading system is to be successful. This error comes about in a similar manner to the previous error, where the researcher has asked a focused question without an eye toward the application. In particular, he has tried to do a precise price prediction [6] [10] [13] [18](unlike the previous error, this one not based on time-series price inputs), without noting that it is not the actually price he truly needs to predict, but instead the price movement. Upon reflection, it should become obvious that even an error of a large magnitude is bearable if the direction of the prediction is accurate.

In order to create a workable predictive system, the target vector should comprise not of the closing prices themselves, but rather of their actual price movements [4]. Even better would be to have the output prediction broken up into vector format, namely direction and magnitude – in this way, the researcher could impose a higher error-penalty on the magnitude prediction, reflecting the intended purpose of the prediction, and ensuring its usefulness. In this way the error function is not fooled by the proximity to the target or by scaling errors that may have crept into the system.

## VII. INAPPROPRIATE MEASURES OF PERFORMANCE

The problem here lies not with the measurements themselves, but rather on the reliance on them for validating the success of a trading system. These techniques often obscure problems in the system design by looking like successful computational intelligence systems by the standard computational measures. This includes graphs of ROC curves, RMS plots and other typical computational measures of performance [7] [9]. If any of the preceding errors had been made, they would not be picked up by the usual performance measures since they only measure the performance of the system based on the given input and output values. This is a dangerous mistake to make, as the system is still harbouring any mistakes made, but the researcher is happy to carry on, secure in the success of his system, verified by an inappropriate measure of performance. Reliance on these measures comes naturally to researchers in this field as they form the benchmark of most computational intelligence and machine learning methods [5], and are thus likely to be utilised almost out of habit by researchers, or often even insisted upon by supervisors.

Instead of the above, the designer should set up a trading simulator, and use the designed predictor to simulate trading based on its predictions. By performing actual trades based upon the predictions, many of the errors described in this paper will be quickly exposed, as the actual trading results will be poor, or at best highly erratic. The nature of the errors will often become apparent when measuring performance in this manner, matching those described within each section, making it a much more useful measure of performance both during and after the system design process.

## VIII. CONCLUSIONS

This paper has examined a number of different errors that are common occurrences in computational intelligence trading literature. The first mistake examined is a development cycle error, in which the development cycle creates a need for additional datasets, and an appropriate calculation for number of datasets was recommended. The second mistake is a pre-processing mistake, on which there is no shortage of literature in order to avoid this sort of mistake. The third mistake is really about vision and expectations, and should be immediately picked up by any researcher with reasonable expectations, and an examination of input-output airs was recommended in order to prevent this type of mistake. The fourth mistake is also a mistake of vision, essentially mis-matching the required variable with a different predicted variable, and a precise example was suggested of a useful prediction for the task at hand. The fourth and final mistake examined is a mistake of assumptions, namely assuming that the researcher's input-ouput formulation has been correct, and that all that matters in order to be certain of the validity of the model is a low error value. A more appropriate measure of performance was recommended, one that evaluates performance at execution rather than at implementation, and thus evaluates the performance outside of the limited scope of the researcher's assumptions.

The errors themselves have been highlighted, their causes and their effects. Alternative approaches that help to circumvent these problems have been shown. Lastly, an alternative method for measuring performance has been proposed, one that allows the designer to immediately detect if the system is malfunctioning and also aids in diagnosing the problem(s) should they appear. It is this researcher's hope that with more reliable research emerging, the potential for these techniques will finally be realised in the marketplace.